\documentclass{article}

\usepackage{arxiv}

\usepackage[utf8]{inputenc} 
\usepackage[T1]{fontenc}    
\usepackage{hyperref}       
\usepackage{url}            
\usepackage{booktabs}       
\usepackage{amsfonts}       
\usepackage{nicefrac}       
\usepackage{microtype}      
\usepackage{lipsum}
\usepackage{amsmath}
\usepackage{graphicx}

\title{The minimum of the time-delay wavefront error in Adaptive Optics}

\author{
  Niek Doelman\thanks{2nd affiliation: Leiden Observatory, Leiden University, The Netherlands.} \\
  Department of Opto-Mechatronics\\
  TNO Industry\\
  Delft, The Netherlands\\
  \texttt{niek.doelman@tno.nl} \\
}
\begin{document}
\maketitle

\begin{abstract}
	An analytical expression is given for the minimum of the time-delay induced
	wavefront error (also known as the servo-lag error) in Adaptive Optics systems under temporal prediction filtering. The analysis is based on the von K\'{a}rm\'{a}n model for the spectral density of refractive index fluctuations and the hypothesis of frozen flow. An optimal, temporal predictor can achieve up to a factor 1.77 more reduction of the wavefront phase variance compared to the zero-order prediction strategy, which is commonly used in Adaptive Optics systems. Alternatively, an optimal predictor can allow for a 1.41 times longer time-delay to arrive at the same residual phase variance. Generally, the performance of the optimal, temporal predictor depends on the very product of time-delay, wind speed and the reciprocal of turbulence outer scale.
\end{abstract}



\section{Introduction}
\label{sec:intro}  
- The residual wavefront error of an astronomical imaging instrument equipped with an Adaptive Optics (AO) system, is determined by several error sources. The instrumental-type errors represent the limitations of the AO system components to cancel the turbulence-induced wavefront distortion. One of the most prominent instrumental AO error sources is the time-delay error, which is due to the overall latency between the sensing and the actual correction of the wavefront. This error is also known as the AO servo-lag error.

In \cite{fried1990time} the impact of the time-delay on the residual AO wavefront error has been described. A specific analytical expression is given for the wavefront error variance, for Kolmogorov turbulence and at a single point.  The time-delay analysis is based on a control approach, of which the essence is to feed the latest (measured) wavefront phase value, with opposite sign, back to the optical wavefield. In continuous-time notation, this control action leads to the residual phase: 
\begin{equation}
	\label{eq:interror}
	\epsilon_{ref}(t) = \phi(t+\Delta \hspace{0.5pt} t) -\phi(t)
\end{equation}
in which $\phi(t)$ represents the wavefront phase fluctuations as a function of time $t$, $\Delta \hspace{0.5pt} t$ the time-delay and $\epsilon_{ref}(t)$ the residual phase.
This control strategy is very common in AO systems and is denoted as \emph{zero-order prediction} in \cite{hardy1998adaptive}. For the remainder of the paper it will be denoted as the reference approach.\par
In practice, many closed-loop AO systems utilise the zero-order prediction strategy, in the form of a discrete-time integrator. It needs to be noted that the integrator, as a result of the closed-loop stability constraint, may not obtain the same performance as the zero-order prediction strategy.\par
The variance of the residual wavefront error for the reference prediction approach (\cite{fried1990time}) is evaluated as 
\begin{equation}
	\label{eq:refvar}
	\sigma^{2}_{ref} = 28.4 (f_G \Delta \hspace{0.5pt} t)^{{5}/{3}}
\end{equation}
with $f_G$ the Greenwood frequency.

This analytical expression is often used in AO performance analysis, design and error budgeting. The underlying reference controller has the advantage of having the lowest possible order and being straightforward to implement. Yet, it does not achieve the minimum possible value of the mean square time-delay error.
\subsection{Optimal prediction}
More advanced prediction methods to reduce the effect of the AO time-delay have been proposed by various researchers. In particular, optimal prediction has gained a lot of attention since it aims at achieving the minimum of the mean square phase error. 
In (\cite{paschall1993linear}) this principle is proposed in the context of AO within a closed-loop Linear Quadratic Gaussian control approach. Since then, optimal prediction for AO has been discussed by many authors; see for instance (\cite{kulcsar2017towards}) for a recent overview and references therein. The large majority of the work on AO predictive control has been focused on numerical simulations, laboratory experiments or on-sky telescope verification tests. In those settings, prediction is performed in discrete-time and in a closed AO control loop. And therefore, the specific temporal response behaviours of deformable mirror (DM) and wavefront sensor (WFS) need to be accounted for. Bandwidth limitations of DM and WFS may degrade the overall AO performance and hence may hold back the potential benefit of optimal prediction. The work of (\cite{harrington1994frequency}) already describes this for the reference prediction approach.
\subsection{Scope}
This paper addresses least-square optimal prediction from an analytical and continuous-time point of view. An analytical expression is derived for the minimum wavefront error variance with an optimal, temporal prediction filter. This predictor is based on present and past wavefront phase values only. The minimum variance expression holds for von K\'{a}rm\'{a}n type optical turbulence. Since it is analytical, it clearly shows the behaviour of the residual variance as a function of the key parameters:  Greenwood frequency, outer scale, wind speed and time-delay.
Since the potentially limiting properties of DM and WFS are not taken into account, the analytical expression can serve as a performance upper bound for the servo-lag error of AO systems in practice. Furthermore, it can be used next to Fried's expression (\ref{eq:refvar}) for the reference predictor to quantify the potential benefit of optimal prediction in particular turbulence cases. This benefit can be either in terms of a lower phase variance or an enhanced detector integration time.
Besides optimal prediction, the paper gives the analytical expression for the reference predictor under von K\'{a}rm\'{a}n optical turbulence. This can be regarded as an extension to Fried's expression (\ref{eq:refvar}), which only applies to Kolmogorov turbulence.
\subsection{Structure of the paper}
In the upcoming sections, a stochastic process model will be derived for the wavefront phase fluctuations based on the von K\'{a}rm\'{a}n spectral density (Section 2). This model follows from a factorization of the von K\'{a}rm\'{a}n power spectrum. Section 3.1 will show that the stochastic model leads to an analytical expression for both the optimal predictor and the minimum mean square value of the time-delay wavefront error. An extension of the Fried expression for the reference predictor under von K\'{a}rm\'{a}n turbulence is given in Section 3.2. The specific case of Kolmogorov turbulence for both predictors is addressed in Section 3.3. Section 4 will analyse the properties of optimal, temporal prediction and relate those to the reference prediction approach. The paper will finish up with an analysis for the 'path-integrated case' (Section 5) and final conclusions.

\section{Stochastic process model}
\label{sec:stochastic_model}
\subsection{Power Spectral Density}
Consider a turbulent atmospheric layer of thickness $\delta h$ at height $h_i$ and an incident plane wave under a zenith angle $\zeta$. The turbulence is assumed to be stationary, homogeneous and isotropic and is described by the von K\'{a}rm\'{a}n model for index-of-refraction fluctuations. 
The wave distortion after propagation through the thin turbulent layer can be characterised by the covariance function of the wavefront phase fluctuations (\cite{andrews2005laser,conan2008mean}) and references therein as:
\begin{equation}
	\label{eq:Covar_vonK}
	C_{\phi}(r) = \frac{\Gamma(\frac{7}{6})}{\sqrt{2}\hspace{1mm}\pi^{\frac{5}{3}} \hspace{1mm} \Gamma(\frac{1}{3})}  k^{2}\delta z C_{n}^2 {\kappa_{0}}^{-\frac{5}{3}}(2\pi \kappa_{0} r)^{\frac{5}{6}} K_{\frac{5}{6}}(2\pi \kappa_{0} r)
\end{equation}
where $\Gamma(\frac{7}{6})/(\sqrt{2} \hspace{1mm} \pi^{\frac{5}{3}} \hspace{1mm} \Gamma(\frac{1}{3})) = 0.0363$, $K_{5/6}$ is the modified Bessel function of the second kind of order $5/6$, $\kappa_{0} =  1/L_0(h_i)$, $L_0(h_i)$ is the outer scale of the atmospheric turbulent layer, $\delta z = \delta h \hspace{1pt} \sec(\zeta) $, $k$ is the wavenumber and $r= \left| \mathbf{r} \right|$. The parameter $C_{n}^2$ represents the index-of-refraction structure constant at $h_i$. \par
The covariance function $C_{\phi}(r)$ above is circularly symmetric. Taylor's hypothesis of frozen flow implies that for a turbulence variable $u(\mathbf{r},t)$ it holds, that the future value at $t+\tau$  can be written as a spatially shifted value at $t$: $u(\mathbf{r},t+\tau) = u(\mathbf{r} - \mathbf{v}\tau,t)$. Under this hypothesis the spatial covariance function can be converted to a temporal covariance function for a single point by replacing the spatial variable $r$ by $v\tau$; $C_{\phi}(\tau) = C_{\phi}(r)$ with $r=v \tau$ . Here the variable $v = \left| \mathbf{v_{\bot}} \right|$ represents the modulus of the wind speed perpendicular to the propagation direction at height ${h_i}$.

The Fourier Transform of the temporal covariance function, $\int_{-\infty}^\infty C_{\phi}(\tau) \exp(- i \omega \tau) d\tau$, renders the power spectral density (PSD) of phase fluctuations (see eq. 6.699/12 in \cite{gradshteyn2007table}):
\begin{equation}
	\label{eq:PSD_temp}
	\Phi(\omega)= \frac{4}{3} \sqrt{\pi} \hspace{1mm} \Gamma(\frac{7}{6}) \hspace{1mm} k^{2}\delta z C_{n}^2 \frac{v^{\frac{5}{3}}}{\left(\omega^2 + \omega_0^2\right)^{\frac{4}{3}}}
\end{equation}
where $\frac{4}{3} \sqrt{\pi} \hspace{1mm} \Gamma(\frac{7}{6})=2.19$ and $\omega_0 = 2 \pi v/L_{0}$. The frequency $\omega_0$ can be regarded as the angular cut-off frequency in the PSD. The function $\Phi(\omega)$ is double-sided and has unit rad$^{2}$/Hz. 
Equation (\ref{eq:PSD_temp}) is also given by (\cite{harrington1994frequency}), who derived the PSD following a different route.

\par
Expression (\ref{eq:PSD_temp}) can be clarified further. 
For the case of a single turbulent layer at $h_i$ the Greenwood frequency $f_{G}$ can be expressed as (\cite{greenwood1977bandwidth}):
\begin{equation}
	\label{eq:Greenwood_i}
	f_{G} = \left[ \frac{2^{\frac{1}{3}} \hspace{1mm} \Gamma(\frac{7}{6})}{3 \hspace{1mm} \pi^{\frac{7}{6}}}\hspace{1mm} k^{2} C_{n}^2(h_i) v^{\frac{5}{3}}(h_i) \delta z \right]^\frac{3}{5}
\end{equation}
where $ {2^{1/3} \hspace{1mm} \Gamma(7/6)}/{3 \hspace{1mm} \pi^{7/6}} = 0.102$.
Inserting $f_{G}$ into the expression for the PSD (\ref{eq:PSD_temp}) gives
\begin{equation}
	\label{eq:PSD_temp2}
	\Phi(\omega) =  \frac{(2 \pi f_{G})^{\frac{5}{3}}}{\left(\omega^2 + \omega_0^2\right)^{\frac{4}{3}}}
\end{equation}
In the limit of an unbounded outer scale $L_0 \rightarrow \infty$ and therefore $\omega_0 \downarrow 0$, the expression for the power spectral density is reduced to
\begin{equation}
	\label{eq:PSD_kolmo}
	\Phi_{Kol}(\omega) =  (2 \pi f_{G})^{\frac{5}{3}}\omega^{-\frac{8}{3}}
\end{equation}
This is in fact the power spectral density for the case of Kolmogorov turbulence and is in full agreement with eq.(11) in \cite{fried1990time} and with the abstract formalism given in (\cite{conan1995wave}). \par
The variance of the uncorrected or primary wavefront phase fluctuations - $C_{\phi}(r)$ for $r=0$ in (\ref{eq:Covar_vonK}) - amounts to:
\begin{equation}
	\label{eq:prim_var}
	\sigma^{2}_{prim} = \frac{3 \hspace{2pt} \Gamma(\frac{5}{6})}{2 \sqrt{\pi} \Gamma(\frac{1}{3})} \left(\frac{f_{G}}{f_0} \right)^{\frac{5}{3}} 
\end{equation}
where  $3 \hspace{1pt} \Gamma(\frac{5}{6})/(2 \sqrt{\pi} \hspace{1pt} \Gamma(\frac{1}{3})) = 0.357$ and $f_0 = \omega_0 /(2\pi) = v/L_0$. The primary variance increases with the $5/3$ power of the $f_G/f_0$ fraction. For Kolmogorov turbulence the variance is unbounded.

\subsection{Spectral factor}
Modeling the wavefront phase fluctuations as a real, wide-sense stationary random process, $\phi(t)$ can be represented in an innovations model form:
\begin{equation}
	\label{eq:innov}
	\phi(t)= \int\limits_0^\infty h(\tau)\xi(t-\tau)d\tau
\end{equation}
where $\xi$ is zero-mean white noise process, with auto-covariance function: $R_{\xi\xi}(\tau)=\delta(\tau)$. The causal innovations filter $h(\tau)$ is the impulse response of transfer function $H(s)$, which is the minimum-phase spectral factor of the power spectrum $\Phi(s)$, such that $\Phi(s) = H(s)H(-s)$, see \cite{papoulis1991probability}. 
By taking the Laplace transform of the covariance function (\ref{eq:Covar_vonK}), $\int_{-\infty}^\infty C_{\phi}(\tau) \exp(-s \tau) d\tau $, the power spectrum $\Phi(s)$ is obtained as
\begin{equation}
	\label{eq:PSD_laplace}
	\Phi(s) =  \frac{(2 \pi f_{G})^{\frac{5}{3}}}{\left(-s^2 + \omega_0^2\right)^{\frac{4}{3}}}
\end{equation}
This power spectrum (\ref{eq:PSD_laplace}) obeys the Paley-Wiener criterion. Its minimum-phase spectral factor can be readily found by taking the left-hand side roots of $\Phi(s)$:
\begin{equation}
	\label{eq:spectralfactor}
	H(s)= \frac{(2 \pi f_{G})^{\frac{5}{6}}}{\left(s+\omega_0\right)^{\frac{4}{3}}}
\end{equation}
The impulse response of the spectral factor (\ref{eq:spectralfactor}) follows by taking the inverse Fourier transform:
\begin{equation}
	\label{eq:impulseresponse}
	h(\tau) = \frac{1}{2 \pi} \int\limits_{-\infty}^\infty H(i \omega) \exp(i \omega \tau) d\omega = \frac{(2 \pi f_{G})^{\frac{5}{6}}}{\Gamma(\frac{4}{3})} \hspace{1mm} \tau^{\frac{1}{3}}\exp(-\omega_0\tau)
\end{equation}
see eq. 3.382/6 in \cite{gradshteyn2007table}). The impulse response (\ref{eq:impulseresponse}) is causal, $h(\tau) = 0$ for $\tau < 0$.

\subsection{Related model families}
The covariance function of wavefront phase fluctuations (\ref{eq:Covar_vonK}) belongs to the family of Mat\'ern functions (\cite{matern1960spatial}). In the particular case of the von K\'{a}rm\'{a}n model, the Mat\'ern smoothness parameter equals $5/6$.\par
Time series models such as damped fractional Brownian motion or tempered fractionally integrated models (\cite{sabzikar2015tempered}) can be regarded as discrete-time versions of Mat\'ern processes and hence can be representative for von K\'{a}rm\'{a}n type phase fluctuations.
In accordance with (\ref{eq:Covar_vonK}), the fractional integration parameter then equals $4/3$ and the tempering parameter is determined by $\omega_0$.\par
Next to existing, numerical methods to simulate von K\'{a}rm\'{a}n type wavefront phase screens - see for example (\cite{assemat2006method}) - the Mat\'ern type models may add further physical and analytical insight into the stochastic behaviour of wavefront phase time series and fields.

\section{Residual time-delay error}
\subsection{Optimal prediction}
\label{sec:minimum}
To determine the minimum of the mean square time-delay error, the optimal prediction of the phase fluctuations $\phi(t)$ needs to be formulated. Given the overall time-delay $\Delta \hspace{0.5pt} t$ of the AO loop, at time instant $t$ the future value $\phi(t+\Delta \hspace{0.5pt} t)$ is to be predicted based on its time history $\phi(t-\tau)$, with $\Delta \hspace{0.5pt} t > 0$ and $\tau\geq 0$. Note that the predictor relies on temporal information only. Denoting the predictor as a causal linear, time-invariant (LTI) filter $P$, the prediction of $\phi(t+\Delta \hspace{0.5pt} t)$ can be expressed as:
\begin{equation}
	\label{eq:estimator}
	\widehat{\phi}(t+\Delta \hspace{0.5pt} t) = \int\limits_0^\infty p(\tau)\phi(t-\tau)d\tau
\end{equation}
Based on the innovations model (\ref{eq:innov}, \ref{eq:impulseresponse}), the optimal prediction filter of $\phi(t)$ over an horizon $\Delta \hspace{0.5pt} t$ can be derived, see \cite{papoulis1991probability}. The Laplace domain optimal predictor equals: 
\begin{equation}
	\label{eq:optPRED1}
	P_{opt}(s) =   \frac{1}{H(s)} \hspace{1pt} \int_{0}^\infty h(\tau + \Delta \hspace{0.5pt} t) \exp(-s \tau) d\tau
\end{equation}
Inserting the spectral factor (\ref{eq:spectralfactor}) and (\ref{eq:impulseresponse}) yields:
\begin{equation}
	\label{eq:optPRED2}
	P_{opt}(s) = \frac{1}{\Gamma(\frac{4}{3})} \hspace{1pt} \exp(s\Delta \hspace{0.5pt} t)\hspace{1pt} \Gamma_{u}(\frac{4}{3}, (s+\omega_0)\Delta \hspace{0.5pt} t)
\end{equation}
where $\Gamma_{u}$ is the upper incomplete gamma function: $\Gamma_u(a,x) = \int_{x}^{\infty} t^{a-1}\exp(-t)dt$.
The corresponding minimum of the mean square time-delay error can be written as (\cite{papoulis1991probability}):
\begin{equation}
	\label{eq:minimum}
	\sigma^{2}_{min} = \int\limits_0^{\Delta \hspace{0.5pt} t} h^{2}(\tau)d\tau
\end{equation}
Using (\ref{eq:impulseresponse}), this minimum mean square error equals:
\begin{equation}
	\label{eq:optMSE}
	\sigma^{2}_{min} = \frac{1}{\Gamma^2(\frac{4}{3})} \left(\frac{f_{G}}{2f_0} \right)^{\frac{5}{3}} \gamma_\ell(\frac{5}{3},2 \omega_0 \Delta \hspace{0.5pt} t)
\end{equation}
where $\gamma_\ell$ is the lower incomplete gamma function: $\gamma_\ell(a,x) = \int_0^{x} t^{a-1}\exp(-t)dt$.\\

\subsection{Zero-order prediction under von K\'{a}rm\'{a}n turbulence}
The main principle of the reference prediction approach in AO systems is to feed the latest (measured) wavefront phase value, with opposite sign, back to the optical wavefield. The resultant phase variance will now be compared to the minimum value (\ref{eq:optMSE}). In the form of the prediction expression (\ref{eq:estimator}), the transfer function of the zero-order predictor is $P_{ref}(s) = 1$, leading to the residual error (\ref{eq:interror}). This is in fact the differenced wavefront phase over an interval $\Delta \hspace{0.5pt} t$. Hence, the von K\'{a}rm\'{a}n structure function of wavefront phase fluctuations  $D_{\phi}(r)$ (at $r=v \Delta \hspace{0.5pt} t$) exactly represents the residual variance of the reference approach. This leads to:

\begin{equation}
	\label{eq:intMSE2}
	\begin{split}
		\sigma^{2}_{ref} &=  D_{\phi}(v \Delta \hspace{0.5pt} t) = 2 \left[\sigma_{prim}^{2} - C_{\phi}(v \Delta \hspace{0.5pt} t) \right] \\
		& = \frac{1}{\sqrt{\pi}\hspace{1mm}\Gamma(\frac{4}{3})}\left(\frac{f_G}{f_0}\right)^{\frac{5}{3}}\left[\Gamma(\frac{5}{6})-2^{\frac{1}{6}}(\omega_0 \Delta \hspace{0.5pt} t)^{\frac{5}{6}} K_{\frac{5}{6}}(\omega_0\Delta \hspace{0.5pt} t) \right]
	\end{split}
\end{equation}
where $K_{5/6}$ is the modified Bessel function of the second kind of order $5/6$. \par

\subsection{Case of Kolmogorov Turbulence}
\label{sec:Kolmogorov}
The expressions for residual phase error variance (\ref{eq:optMSE}) and (\ref{eq:intMSE2}) hold for any non-negative value of the time-delay $\Delta \hspace{0.5pt} t$. In practice, the delay will be limited and the product $\omega_{0}\Delta \hspace{0.5pt} t$ will be much smaller than unity, even for a high wind speed and a small outer scale. Evaluation of a series expansion of (\ref{eq:optMSE}) and (\ref{eq:intMSE2}) for small  $\omega_0 \Delta \hspace{0.5pt} t$ leads to:
\begin{align}
	\label{eq:optMSE_approx3}
	\sigma^{2}_{min} \approx \left(2 \pi f_G \Delta \hspace{0.5pt} t\right)^{5/3}\left[0.753-0.941 (\omega_{0}\Delta \hspace{0.5pt} t) + \mathcal{O}(\omega_{0}\Delta \hspace{0.5pt} t)^{2}\right]\\
	\sigma^{2}_{ref} \approx \left(2 \pi f_G \Delta \hspace{0.5pt} t\right)^{5/3}\left[1.33-1.07 (\omega_{0}\Delta \hspace{0.5pt} t)^{\frac{1}{3}} + \mathcal{O}(\omega_{0}\Delta \hspace{0.5pt} t)^{2}\right]
\end{align}
For the specific case of Kolmogorov turbulence, in the limit of $L_0 \rightarrow \infty$ and so $f_0 \downarrow 0$, the expressions for the residual variances reduce to:
\begin{align}
	\label{eq:minMSElimit}
	\lim_{f_0\downarrow 0} \sigma^{2}_{min} = \frac{3}{5 \hspace{1mm} \Gamma^2(\frac{4}{3})} (2\pi f_G \Delta \hspace{0.5pt} t)^{\frac{5}{3}} \approx 16.1 (f_G \Delta \hspace{0.5pt} t)^{\frac{5}{3}}\\
	\label{eq:intMSElimit}
	\lim_{f_0\downarrow 0} \sigma^{2}_{ref} = \frac{3}{5} \frac{\Gamma(\frac{1}{6})}{2^{\frac{2}{3}} \sqrt{\pi} \hspace{1pt} \Gamma(\frac{4}{3})} (2\pi f_G \Delta \hspace{0.5pt} t)^{\frac{5}{3}} \approx 28.4 (f_G \Delta \hspace{0.5pt} t)^{\frac{5}{3}} 
\end{align}
Note that the expression for $\sigma^{2}_{ref}$ is equal to eq. (20) in \cite{fried1990time}.\\
So, both residual variances increase with the $5/3$ power of the product $f_G\Delta \hspace{0.5pt} t$. The minimum wavefront phase error with the optimal predictor is a factor $[\Gamma(\frac{4}{3})\Gamma(\frac{1}{6})]/[2^{\frac{2}{3}}\sqrt{\pi}] = 1.77$ smaller than with the reference approach.\par

\section{Analysis}
\label{sec:Analysis}
The optimal predictor (\ref{eq:optPRED2}) is a function of time-delay $\Delta \hspace{0.5pt} t$, wind speed $v$ and outer scale $L_0$. It does not depend on for instance the wavenumber $k$, zenith angle $\zeta$ or the index-of-refraction structure constant $C_{n}^2$.\par
From (\ref{eq:prim_var}), (\ref{eq:optMSE}) and (\ref{eq:intMSE2}) it follows that the minimum of the mean square time-delay error increases with $\Delta \hspace{0.5pt} t$ and decreases with $f_0$. Similar to the primary variance and the residual phase variance with the reference approach, the minimum variance grows with the ${5/3}$ power of the Greenwood frequency. In addition, the normalised residual variance ($\sigma^{2}/\sigma_{prim}^{2}$) is a function of the product $f_0 \Delta \hspace{0.5pt} t$, for both the optimal predictor and the reference predictor.\par
\begin{figure}
	\centering
	\mbox{\includegraphics[width=0.7\linewidth]{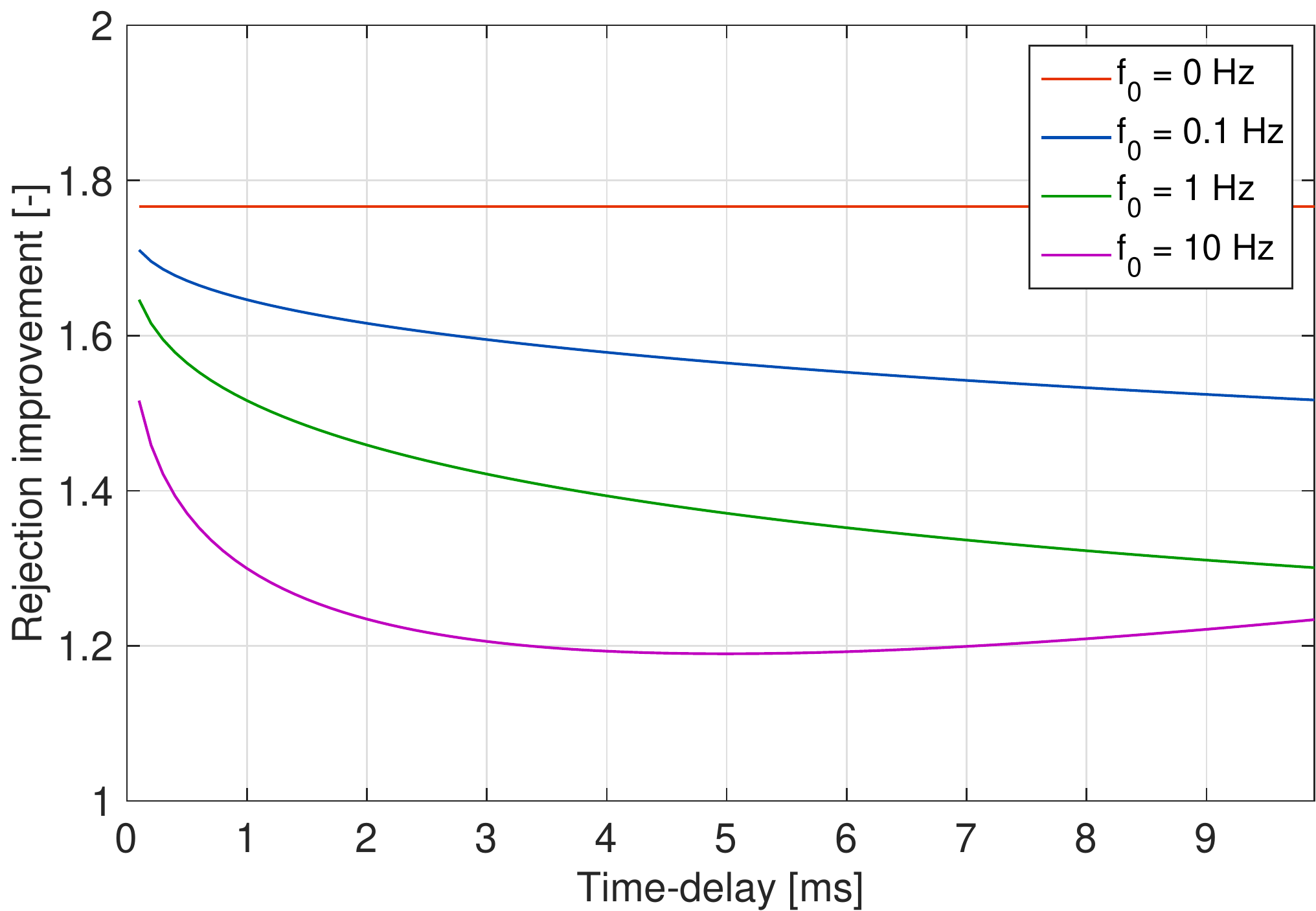}}
	\caption{Improvement on variance reduction of predictor versus reference ($\sigma^{2}_{ref}/\sigma^{2}_{min}$) for small time-delays and various values of $f_0$.}
	\label{fig:Improv}
\end{figure}
\begin{figure}
	\centering
	\mbox{\includegraphics[width=0.7\linewidth]{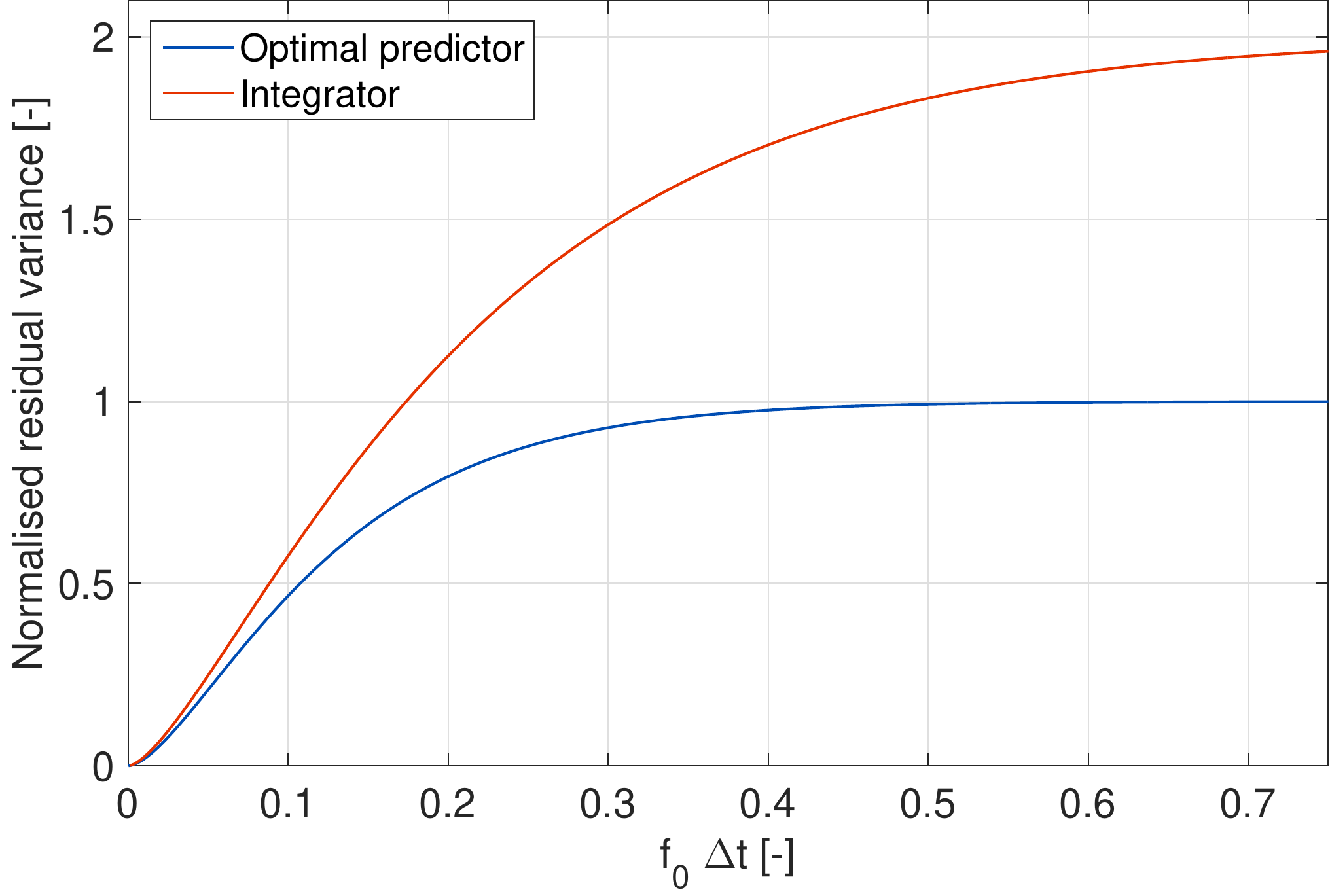}}
	\caption{Normalised residual phase variances $\sigma^{2}_{min}/\sigma^{2}_{prim}$ and $\sigma^{2}_{ref}/\sigma^{2}_{prim}$ against $f_0 \Delta \hspace{0.5pt} t$.}
	\label{fig:ResidVarlong}
\end{figure}

The optimal predictor always performs better than the reference; see Figures \ref{fig:Improv}, and \ref{fig:ResidVarlong}. For small values of $f_0\Delta \hspace{0.5pt} t$, the improvement on phase variance reduction ranges from a factor 1.19 (for $f_0\Delta \hspace{0.5pt} t = 0.05$) up to 1.77 for $f_0\Delta \hspace{0.5pt} t = 0$, which represents the Kolmogorov turbulence case.  For large $f_0 \Delta \hspace{0.5pt} t$, the optimal predictor becomes ineffective and achieves no phase error reduction (for $f_0\Delta \hspace{0.5pt} t > 0.4$). For the reference predictor, a large $f_0\Delta \hspace{0.5pt} t$ value ($> 0.75$) leads to doubling of the primary variance, as phase disturbance values large $\Delta \hspace{0.5pt} t$ apart are fully uncorrelated; see Figure \ref{fig:ResidVarlong}. 
Note, that these large $f_0\Delta \hspace{0.5pt} t$ values are unlikely in practical cases.\par
Apart from a smaller temporal wavefront error, another benefit of optimal prediction in AO systems would be to allow for a longer detector integration time and therefore the use of a fainter reference star.  This was proposed in for instance (\cite{doelman2004}) and discussed in further detail in (\cite{Hinnen2005}). With the analytical expressions for the residual variance, (\ref{eq:minMSElimit}) and (\ref{eq:intMSElimit}) in the Kolmogorov case, the exact gain in integration time can now be quantified. Equalling the two variances gives the following relation between the delay times: 

\begin{equation}
	\label{eq:delaytimes}
	\Delta \hspace{0.5pt} t_{opt} = \left[\frac{\Gamma(\frac{4}{3})\Gamma(\frac{1}{6})}{2^{\frac{2}{3}}\sqrt{\pi}}\right]^{\frac{3}{5}} \Delta \hspace{0.5pt} t_{ref} \approx 1.41  \Delta \hspace{0.5pt} t_{ref}
\end{equation}
So the detector integration time with an optimal predictor can be 1.41 times longer compared to the reference case. This would allow a higher reference star magnitude and would improve the sky coverage.\par
\begin{figure}
	\centering
	\mbox{\includegraphics[width=0.7\linewidth]{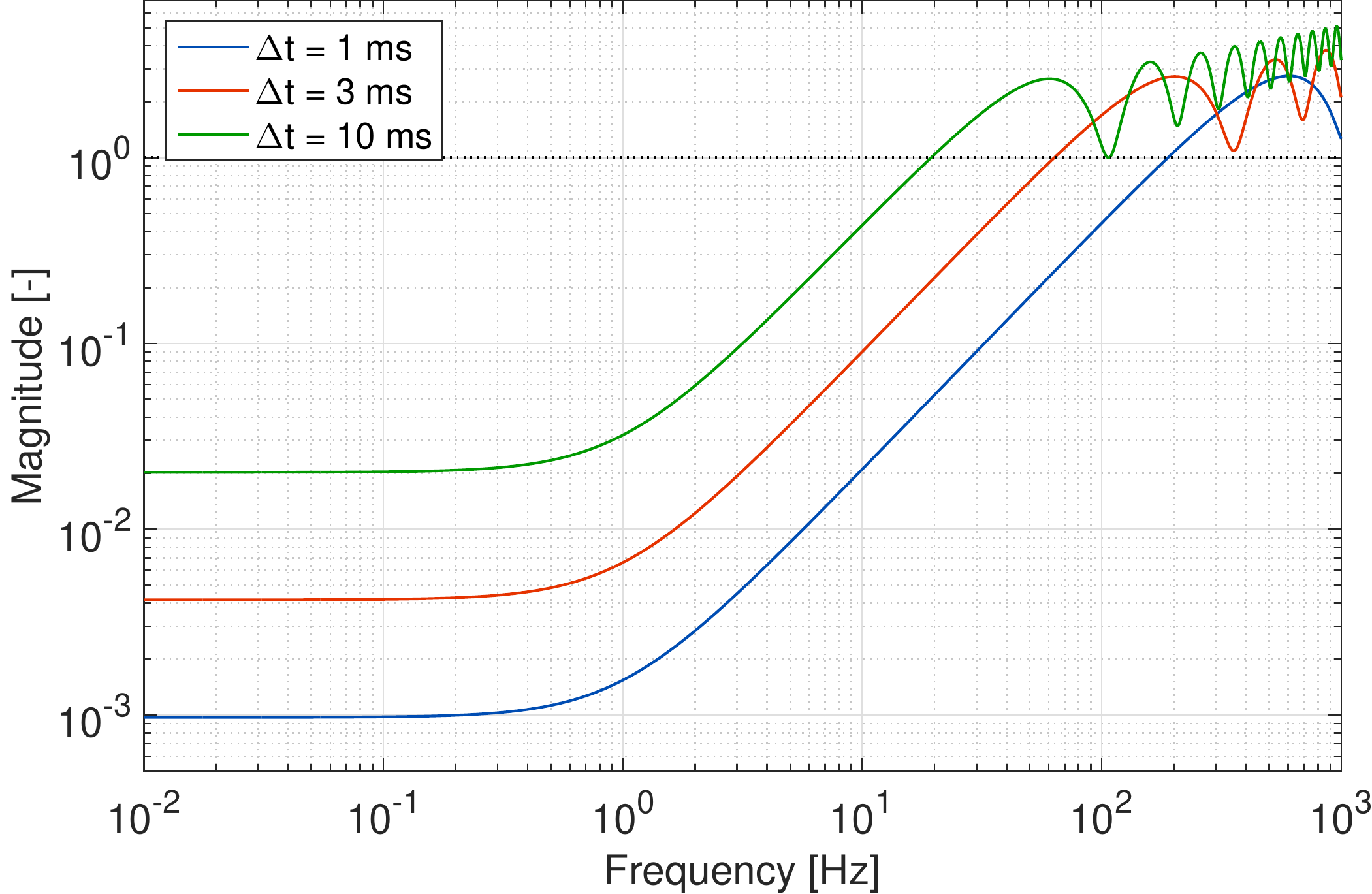}}
	\caption{Optimal predictor sensitivity function for $f_0$ = 1 Hz and various values of $\Delta \hspace{0.5pt} t$.}
	\label{fig:Sens_f0}
\end{figure}
\begin{figure}
	\centering
	\mbox{\includegraphics[width=0.7\linewidth]{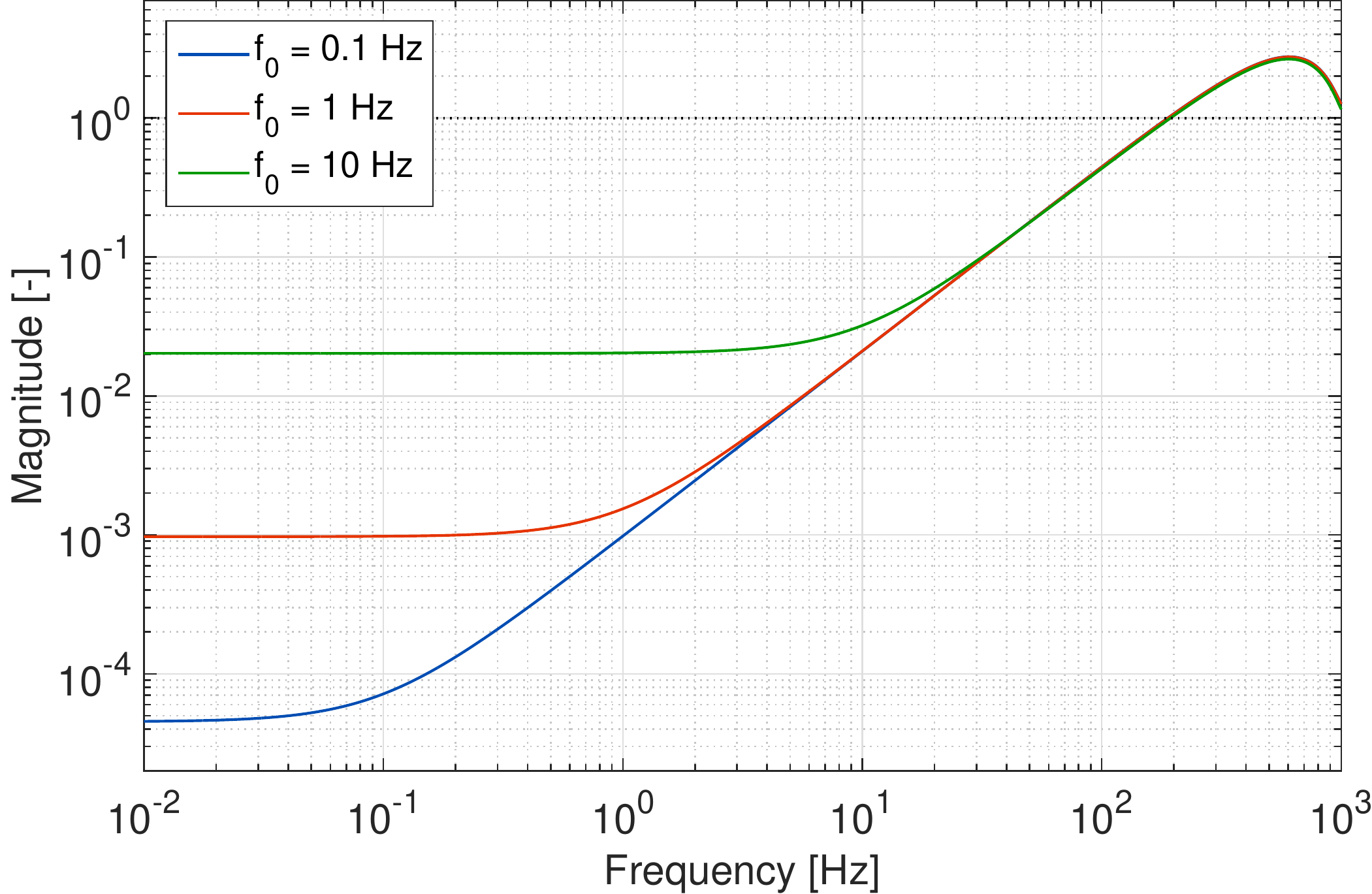}}
	\caption{Optimal predictor sensitivity function for $\Delta \hspace{0.5pt} t$ = 1 ms and various values of $f_0$.}
	\label{fig:Sens_dT}
\end{figure}
\begin{figure}
	\centering
	\mbox{\includegraphics[width=0.7\linewidth]{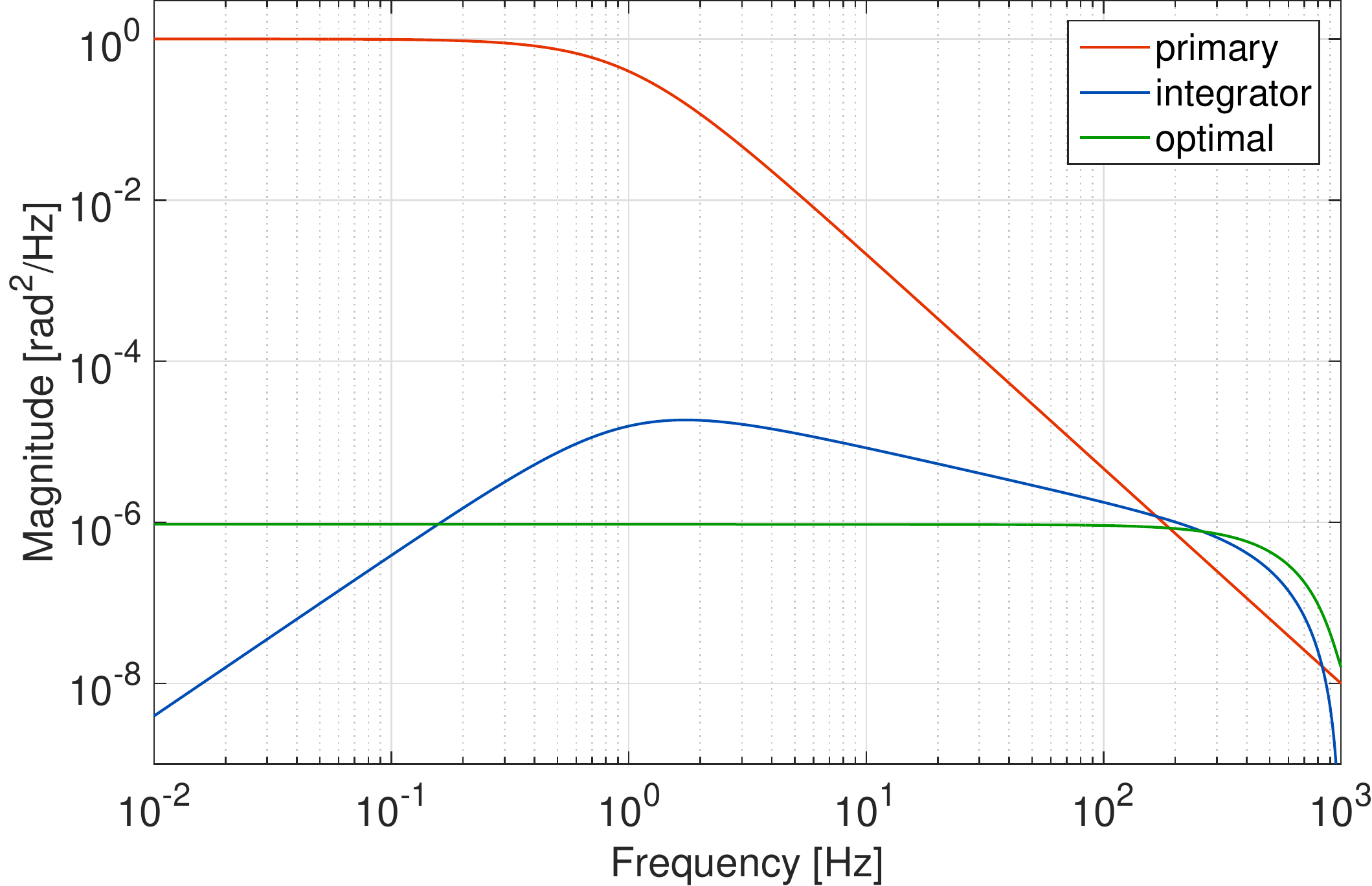}}
	\caption{Power Spectral Density of phase fluctuations for the uncorrected case, the reference and the optimal predictor for $\Delta \hspace{0.5pt} t$ = 1 ms and $f_0$ = 1 Hz.}
	\label{fig:Sens_comp}
\end{figure}
The spectral behaviour of optimal prediction is shown in Figures \ref{fig:Sens_f0} and \ref{fig:Sens_dT}, which reveal the modulus of the transfer function from input phase error to residual error (i.e. the sensitivity function). Figure \ref{fig:Sens_f0} shows that the bandwidth of rejection reduces with the time-delay. The sensitivity curve crosses the unity magnitude line at approximately $(1/5\Delta \hspace{0.5pt} t)$ Hz. For a fixed time-delay, the bandwidth of rejection is independent of the cut-off frequency $f_0$. Only the degree of low-frequency attenuation is affected by $f_0$; see Figure \ref{fig:Sens_dT}. In terms of the PSD of residual phase fluctuations the optimal predictor achieves a flat spectrum over a large frequency band. It outperforms the reference in the mid-frequency range, whereas the reference obtains a higher rejection in the very low-frequency range; see Figure \ref{fig:Sens_comp}. Both approaches have about the same frequency bandwidth of rejection and give rise to an increase of the high-frequency phase error.

\section{Path-integrated turbulence}
So far the analysis of residual wavefront errors has been restricted to propagation through a single, thin layer of turbulence. The atmosphere for the overall propagation path can be viewed as built up from multiple turbulent layers at different heights. For a plane wave and under the geometrical optics approximation, the path-integrated power spectrum of wavefront phase fluctuations $\overline{\Phi}(\omega)$ can then be written as (\cite{andrews2005laser}):
\begin{equation}
	\label{eq:PSD_integrated}
	\overline{\Phi}(\omega) =\frac{4}{3} \sqrt{\pi} \hspace{1mm} \Gamma(\frac{7}{6}) \hspace{1mm} k^{2} \int\limits_{L}^{} dz \hspace{1mm} C_{n}^2(z) \frac{v^{\frac{5}{3}}(z)}{\left(\omega^2 + \omega_0^2(z)\right)^{\frac{4}{3}}}
\end{equation}
in which $L$ is the propagation path. Here, the PSD cut-off frequency $\omega_0$ is a function of height, since both the outer scale $L_0$ and the wind vector $\mathbf{v}$ are height-dependent. In fact, each turbulent layer has its own, specific PSD cut-off frequency. This prevents getting a similar compact form for the overall power spectrum as in (\ref{eq:PSD_temp}) for a single layer. This phenomenon was already discussed in detail in (\cite{conan1995wave}).\\
Following the approach proposed by several authors (\cite{conan2008mean, andrews2005laser, lukin1999efficient}), an effective cut-off frequency $\overline{\omega_0}$ can be used instead. The $\overline{\omega_0}$ value is then set such as to minimise the discrepancy between the true and approximated power spectrum for instance. This metric can be quantified as:
\begin{equation}
	\label{eq:PSD_error}
	\epsilon(\omega_c) =\int\limits_{-\infty}^{\infty} d \omega \left[ \overline{\Phi}(\omega, \omega_c)  \hspace{2mm} - \hspace{2mm} \overline{\Phi}(\omega) \right]^2
\end{equation}
where $\overline{\Phi}(\omega, \omega_c)$ is $\overline{\Phi}(\omega)$ of eq. (\ref{eq:PSD_integrated}) with $\omega_0(z)$ replaced by the constant $\omega_c$. The optimal value of $\omega_c$ for which $\epsilon(\omega_c)$ is minimised is denoted as $\overline{\omega_0}$.
With the optimal value of the effective cut-off frequency in place, the path-integrated wavefront phase (\ref{eq:PSD_integrated}) PSD becomes:
\begin{equation}
	\label{eq:PSD_temp3}
	\overline{\Phi}(\omega) =  \frac{(2 \pi \overline{f_{G}})^{\frac{5}{3}}} {\left(\omega^2 + \overline{\omega_0}^2\right)^{\frac{4}{3}}}
\end{equation}
in which $\overline{f_G}$ is the Greenwood frequency for the integrated path. It can be evaluated by: $(\overline{f_{G}})^\frac{5}{3}=  \int\limits_{L}^{} f_{G}^{\frac{5}{3}}(z) dz$.\normalcolor

Now, the power spectrum (\ref{eq:PSD_temp3}) has exactly the same form as for the single layer case (\ref{eq:PSD_temp2}). And therefore, the full performance analysis of optimal and reference prediction (section \ref{sec:Analysis}) also holds for the approximated path-integrated case with multiple turbulent layers. Note that for Kolmogorov turbulence specifically, the results of section \ref{sec:Analysis} are still exact as then the cut-off frequency $\omega_0$ plays no role.
\section{Conclusion}
Analytical expressions for the minimum time-delay induced wavefront phase error and the optimal predictor have been presented, under temporal prediction filtering. The specific performance and spectral properties have been analysed. In comparison to the reference zero-order predictor, the performance gain of optimal prediction can be significant. In more detail, the gain depends on the very product of time-delay, wind speed and the reciprocal of outer scale $v \Delta \hspace{0.5pt} t/L_0$. The largest performance advantage is obtained for small values of $v \Delta \hspace{0.5pt} t/L_0$ ($< 10^{-3}$). For larger values -- up to $v \Delta \hspace{0.5pt} t/L_0 = 0.2$ -- the performance gain is modest. \par
The optimal performance results can be viewed as an upper bound for practical, discrete-time implementations of predictive control in AO systems.

\bibliographystyle{unsrt}  

\bibliography{Biblist_Minimum.bib} 

\begin{thebibliography}{10}

\bibitem{fried1990time}
David~L Fried.
\newblock Time-delay-induced mean-square error in adaptive optics.
\newblock {\em JOSA A}, 7(7):1224--1225, 1990.

\bibitem{hardy1998adaptive}
John~W Hardy.
\newblock {\em Adaptive optics for astronomical telescopes}, volume~16.
\newblock Oxford University Press, 1998.

\bibitem{paschall1993linear}
Randall~N Paschall and David~J Anderson.
\newblock Linear quadratic gaussian control of a deformable mirror adaptive
  optics system with time-delayed measurements.
\newblock {\em Applied optics}, 32(31):6347--6358, 1993.

\bibitem{kulcsar2017towards}
Caroline Kulcs{\'a}r, H.-F Raynaud, Jean-Marc Conan, R{\'e}my Juv{\'e}nal, and
  Carlos Correia.
\newblock Towards minimum-variance control of {ELT}s {AO} systems.
\newblock In {\em Proceedings of AO4ELT5 Conference}, 2017.

\bibitem{harrington1994frequency}
Patrick~M Harrington and Byron~M Welsh.
\newblock Frequency-domain analysis of an adaptive optical system's temporal
  response.
\newblock {\em Optical Engineering}, 33(7):2336--2343, 1994.

\bibitem{andrews2005laser}
Larry~C Andrews and Ronald~L Phillips.
\newblock {\em Laser beam propagation through random media}, volume 152.
\newblock SPIE press Bellingham, WA, 2005.

\bibitem{conan2008mean}
Rodolphe Conan.
\newblock Mean-square residual error of a wavefront after propagation through
  atmospheric turbulence and after correction with zernike polynomials.
\newblock {\em JOSA A}, 25(2):526--536, 2008.

\bibitem{gradshteyn2007table}
Izrail~Solomonovich Gradshteyn and Iosif~Moiseevich Ryzhik.
\newblock {\em Table of integrals, series, and products}.
\newblock Academic press, Cambridge, MA, 2007.

\bibitem{greenwood1977bandwidth}
Darryl~P Greenwood.
\newblock Bandwidth specification for adaptive optics systems.
\newblock {\em JOSA}, 67(3):390--393, 1977.

\bibitem{conan1995wave}
Jean-Marc Conan, G{\'e}rard Rousset, and Pierre-Yves Madec.
\newblock Wave-front temporal spectra in high-resolution imaging through
  turbulence.
\newblock {\em JOSA A}, 12(7):1559--1570, 1995.

\bibitem{papoulis1991probability}
Athanasios Papoulis and S~Unnikrishna Pillai.
\newblock {\em Probability, random variables, and stochastic processes}.
\newblock Mc-Graw Hill, 1991.

\bibitem{matern1960spatial}
Bertil Mat{\'e}rn.
\newblock {\em Spatial Variation: Stochastic Models and Their Application to
  Some Problems in Forest Surveys and Other Sampling Investigations}.
\newblock Statens skogsforskningsinstitut, 1960.

\bibitem{sabzikar2015tempered}
Farzad Sabzikar, Mark~M Meerschaert, and Jinghua Chen.
\newblock Tempered fractional calculus.
\newblock {\em Journal of Computational Physics}, 293:14--28, 2015.

\bibitem{assemat2006method}
Fran{\c{c}}ois Ass{\'e}mat, Richard~W Wilson, and Eric Gendron.
\newblock Method for simulating infinitely long and non stationary phase
  screens with optimized memory storage.
\newblock {\em Optics express}, 14(3):988--999, 2006.

\bibitem{doelman2004}
Niek~J. Doelman, Karel J.~G. Hinnen, Freek J.~G. Stoffelen, and Michel~H.G.
  Verhaegen.
\newblock {Optimal control strategy to reduce the temporal wavefront error in
  AO systems}.
\newblock In Domenico~Bonaccini Calia, Brent~L. Ellerbroek, and Roberto
  Ragazzoni, editors, {\em Advancements in Adaptive Optics}, volume 5490, pages
  1426 -- 1437. International Society for Optics and Photonics, SPIE, 2004.

\bibitem{Hinnen2005}
Karel Hinnen, Niek Doelman, and Michel Verhaegen.
\newblock {H2 -optimal control of an adaptive optics system: Part II,
  closed-loop controller design}.
\newblock In Robert~K. Tyson and Michael Lloyd-Hart, editors, {\em Astronomical
  Adaptive Optics Systems and Applications II}, volume 5903, pages 86 -- 99.
  International Society for Optics and Photonics, SPIE, 2005.

\bibitem{lukin1999efficient}
VP~Lukin, EV~Nosov, and BV~Fortes.
\newblock The efficient outer scale of atmospheric turbulence.
\newblock In {\em European Southern Observatory Conference and Workshop
  Proceedings, Astronomy with adaptive optics: present results and future
  programs, ESO/OSA topical meeting, Sonthofen, Germany}, volume~56, page 619,
  1999.

\end{thebibliography}

\end{document}